# Employing WiFi Direct to Build a Wireless Network over both 2.4 GHz and 5.8 GHz bands

Sabur Hassan Baidya  
University of Texas at Dallas  
sabur@utdallas.edu  

Pramod Shirol  
University of Texas at Dallas  
pramod.shirol@utdallas.edu  

Abhishek Basu  
University of Texas at Dallas  
axb101920@utdallas.edu  

Ravi Prakash  
Faculty of Computer Science  
University of Texas at Dallas  
ravip@utdallas.edu  

Abstract

Almost all the WiFi networks today provide single band (either 2.4 GHz or 5.8 GHz) wireless communication functionality for connected mobile nodes. In a single band network, the interference depends on number of nodes in the network and the presence of other networks in the proximity. As the number of nodes in a Network increases, the interference in the network also increases which reduces the throughput of the network. If there are two single band networks, one operating in 2.4 GHz frequency band and other in 5.8 GHz frequency band, then nodes in network 1 will not cause any interference with nodes in network 2. This can be used as a basic idea to implement a network where the nodes in the same network use both the frequency bands to minimize the interference between nodes. We show that building dual band multi-hop network gives better performance in terms of throughput compared to that of a network with single frequency band.

## 1 Introduction

Over the last few years, wireless technology has created a revolution in the field of communication. Since last decade, due to the proliferation of wireless technology, the usage of devices like smart phones, tablets, laptops, wireless enabled cameras and printers has become common practice. As the use of mobile devices is increasing, point-to-point communication between devices and instant access for a service has become an important issue. At present, the most commonly used standard for wireless communication is IEEE 802.11 (WiFi). Using this standard of wireless communication in infrastructure mode, a mobile device communicates with other devices by connecting to an access point (AP). But when two devices need to communicate, connecting through an AP every time will increase the load on the AP. Also, there may be some scenario where one of the devices cannot connect to the AP due to some security restrictions. Because of the aforementioned reasons, the need for direct communication between mobile devices has become inevitable. With the advancement in WiFi direct technology it has become possible for two mobile devices to communicate directly without having to go through a hardware access point.

The present IEEE 802.11 standard uses the 2.4 GHz frequency band (though it supports 5.8 GHz band also) for most of the communications. In the infrastructure mode, all the nodes talk to each other via an access point. But, the number of non-interfering channels are limited in 2.4 GHz frequency band, which may cause interference between nodes when they try to communicate through an AP simultaneously. This has put to a limitation in throughput when the number of simultaneous access to an AP increases. But with the advent of WiFi direct, devices are able to create instant peer-to-peer networks and also maintain the infrastructure connection with Internet. If there are mobile devices with dual WiFi cards connected in a network, these devices can be used to act as software APs to relay data from other devices using WiFi direct. Simultaneously, they can communicate with the infrastructure AP for other works. If this parallel communication uses a combination of two different frequency bands, it can reduce the interference substantially during high traffic demands. As a result, it can increase the throughput of the entire network.

In this paper, we present an efficient way of getting optimal network throughput using WiFi direct based multi-hop dual-bandwidth communication. Wireless nodes with dual bandwidth (2.4 GHz and 5.8 GHz) support, can act as access points to relay data to/from other devices, by emulating a software access point. Any node with dual bandwidth support, after joining the network, will start acting as software AP by advertising the same BSSID of the network that it is connected to. When a new node wants to join the network, it will not differentiate between hardware access point and software access point. New node will find the best possible AP to connect to, based on signal quality and data traffic. When a software AP moves out of range from the connected node or another AP with lower traffic is available, the node will transparently connect to a different AP (hardware or software). We present an algorithm with the combination of 2.4 GHz and 5.8 GHz links to find the best AP (software or hardware) to connect to. In order to find the best AP, we measure the link quality of the paths from the node to the available APs and choose the AP with the best link quality. To measure the link quality, we consider different metrics such as load on access points, signal strength, link capacity and bandwidths supported by access points. We show that, in high traffic conditions, the efficient use of a combination of 2.4 GHz and 5.8 GHz band links can substantially increase the throughput of the entire network as a whole.

The remainder of this paper is organized as follows. Section 2 presents the literature survey for the related works on the point-to-point wireless communication. Section 3 describes the problem formula-tion. Section 4 explores the methods and metrics for performance evaluation. In section 5, we present the solution description including the design and implementation. Section 6 describes the performance evaluation in terms of experimental results and comparison with benchmark. Finally, section 7 concludes the paper.

## 2 Literature Survey

Peer to peer (ad-hoc) wireless communication has been existing with different access technologies starting from infrared in the recent past, to Bluetooth to WiFi at present. However, due to its large range and compatibility with the present Internet architecture, WiFi has become the most dominant technology for p2p communication. Since last few years, researchers came up with different approaches for wireless p2p communication such as wireless ad-hoc network [14], [10] which is infrastructure-less and works for internal communications mostly. However, the wireless ad-hoc network needed an extra support to connect to the infrastructure to provision p2p communication and Internet connection simultaneously. Carlo Parata et al. came up with a solution for that with the implementation of Flex-WiFi [12] in the Madwifi open source driver of Atheros chipset. Flex-WiFi enabled a client node to associate with an access point (AP) and also connect to an ad-hoc network simultaneously but on different physical channels. But the ad-hoc network still does not support multihop connectivity among the nodes which could share the overall network load. To enable this capability of infrastructure based multi-hop p2P communication, wireless mesh network [1] standard (IEEE 802.11s) was evolved as a solution.

However, over the years, as traffic demand has increased, QoS in wireless mesh network has become an important issue of research. Some idea of using OFDMA (Orthogonal Frequency Division Multiple Access) has been implemented [11] to reduce the transmission power and increase fairness among the subcarriers. But, this approach did not increase the overall network capacity. To increase the overall throughput, an architecture and algorithm (called Hyacinth) was proposed [13] by Raniwala et al. for an IEEE 802.11-based multi-channel wireless mesh network. This approach uses multiple non-interfering channels simultaneously for the ad-hoc communication among the nodes having multiple WiFi cards. They devised an algorithm based on link quality metric for dynamic channel assignment among the nodes. They showed that using 2 NICs (Network Interface Cards) per node and applying their algorithm can enhance the throughput by a factor of 6 to 7. Similar algorithms have been proposed in [2] and [8] for using multiple radios and orthogonal channels to improve network throughput. However, they only consider the non-interfering channel assignments among the WiFi nodes from a single frequency band. We are proposing to use a dual band WiFi mesh network, where we can use the links of mixed frequencies (2.4 GHz and 5.8 GHz) to minimize the interference in 2.4 GHz band from other WiFi nodes. This also reduces the interference from other access technologies such as Bluetooth, Microwave which use 2.4 GHz frequency band from interfering with the communication of WiFi network.

The basic model to implement the dual-band wireless mesh network is similar to the implementation of WiFi direct [3], [4], [5]. In WiFi direct, the client nodes scan for other WiFi direct enabled devices and then select a discovered device for connection. However, the device selected to be connected to might not be the best AP in terms of signal strength and link quality. Also, during a communication session, if the existing link quality gets deteriorated due to mobility or other reasons, it needs to shift to other available APs by a smooth handoff. An efficient MAC layer handoff is proposed by Zhang et al. in [16]. According to their approach, when a node decides to connect to a new AP, instead of scanning all the channels by itself, it divides the channels into different groups and assigns each group to a neighboring assistant node to scan the channels for available APs. Decision to do a handoff is made based only on RSSI (Received Signal Strength Indicator) values of the available APs. In our solution, we use a similar technique for handoff but along with signal strength we also consider traffic and bandwidth supported by neighboring nodes for doing handoff.

So, in our solution, we build a wireless network employing WiFi Direct with a combination of 2.4 GHz and 5.8 GHz links and show that during high traffic demand it will enhance the throughput in comparison to the previous related works. Also, unlike most of the earlier works which did the performance evaluation on different simulators, we test our solution with experiments on wireless Testbed which can emulate scenarios closer to the real world.

## 3  Problem Formulation

The basic system architecture this work aims to implement on is multi-hop wireless mesh network. All the nodes in the network will form a wireless mesh and can relay the traffic from one node to other. Some of the nodes in this network act as wireless routers that can connect to the backbone network to provide Internet services in addition to P2P services. To create a dual band network, each node which acts as an access point should be equipped with two 802.11-compliant NICs (Network Interface Cards). One of the NICs supports 2.4 GHz frequency band and other supports 5.8 GHz band. The main goal of this work is to maximize the throughput of the wireless mesh network using a combination of 2.4 GHz and 5.8 GHz links in high traffic conditions. For high traffic scenarios, we consider some applications with large data transfer or applications with high data rate like audio or video streaming. We also need to consider interference from other WiFi devices like Bluetooth as well as non-WiFi devices like Zigbee which can interfere with the communication in 2.4 GHz frequency-band.

Now the underlying problem that we are dealing with in this paper can be segregated into following sub-problems: sensing the link quality based on appropriate metric, developing a protocol for selecting a combination of dual band links, reconfiguring the topology using smooth handoff during high traffic demands and defining performance evaluation metric.

## Metric for Link Quality

RSSI value of the signal can be used as a basic criterion to check the accessibility of an AP from a client. But, in our solution a node acting as a software AP might be having more load on a network card with better RSSI value. For example, there might be more number of nodes connected to a software AP on 2.4 GHz compared to the number of nodes connected on 5.8 GHz. Even if RSSI value of 2.4 GHz is better than that of 5.8 GHz, while joining a new node to the AP, connecting to AP on 5.8 GHz might yield better performance. So, along with signal strength we need to consider the link capacity and delay when the clients sense an AP.

## Algorithm to create a Dual Band Wireless Network

The closest node acting as a software AP or a hardware AP may not be the best node to be connected to, as different nodes will have different bandwidth and data traffic. But to maximize the throughput, a node needs to connect to the best AP available at that moment. During high traffic conditions, the number of non-interfering links is insufficient in single frequency (2.4 GHz) band to carry the traffic. So, if we can make use of different frequency bands (figure 1) for uplink and downlink, a node can send and receive data simultaneously without much of interference during high traffic conditions. So, we propose an algorithm which can build a dual band (2.4 GHz & 5.8 GHz) wireless network, a dynamic handoff mechanism which can seamlessly rearrange the network topology to maintain the persistence of the network during any dynamism of the network.

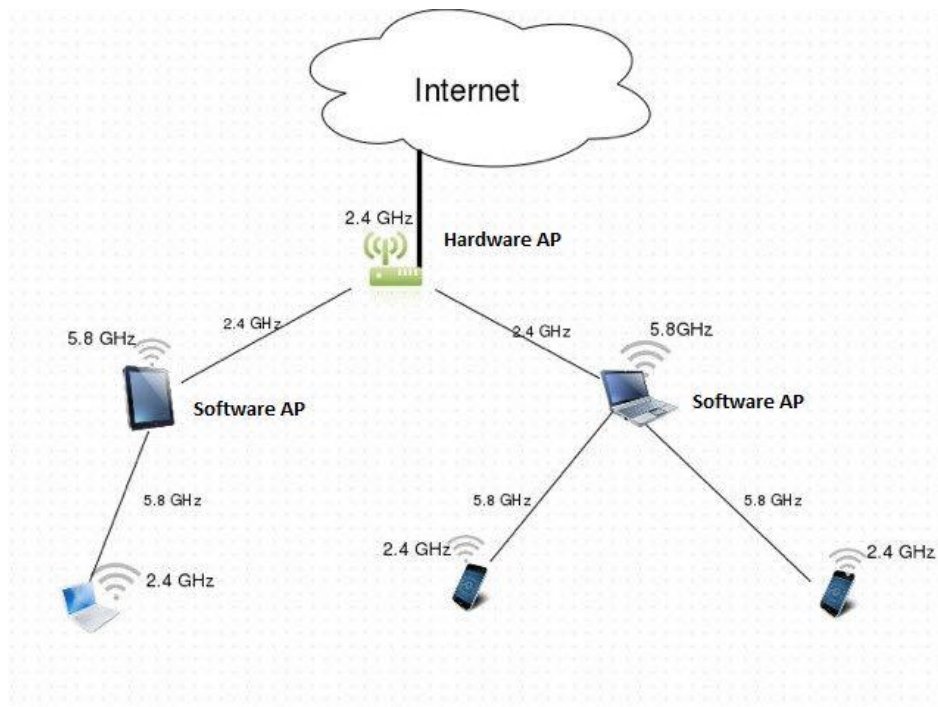

Figure 1: Building a dual band wireless network

### Dynamic Handoff

Changes in network size, traffic demands, node mobility and fluctuations in link quality demand for a change in network topology to improve the overall throughput. But the changes in network topology should not affect the applications running on nodes. This makes dynamic handoff an important issue to be considered. We provide with an algorithm similar to 802.11 handoff algorithm to do MAC layer handoff dynamically when nodes in the network need to be connected to different APs. We ensure that the MAC layer handoff will be transparent and it will not have any impact on applications running on nodes.

### Performance Evaluation Metric

We emulate the environment of real WiFi network to measure the performance of our solution. The network throughput is measured by calculating the average data transfer rate of each node in high traffic. We will compare the throughput obtained from our solution against the throughput of single band network for different topologies and different traffic demands. We will show that multi-hop multi-band network will provide significant improvement compared to single band networks. Detailed description on the methods and metrics for performance evaluation is given in the next section.

## 4 Methods and Metrics for Performance Evaluation

The main objective of building a dual band (2.4 GHz and 5.8 GHz) wireless network employing WiFi Direct is to maximize the throughput of the wireless network in high traffic conditions. The throughput can be measured as the average bit rate of successfully transmitted packets that can be achieved over some duration in the network as a whole.

$$Throughput_{avg} = \frac{\sum_{N \, nodes} Data \ bits \ received \ in \ T \ sec}{T * N} \ bits/sec \quad (1)$$

The comparison benchmark for this work is the maximum achievable throughput in a single band (2.4 GHz) wireless network with WiFi Direct in the same high traffic condition. We are choosing throughput as a performance metric which depends on several other factors whose relations with throughput are linear or non-linear. For a wireless network with WiFi direct, throughput is a function of following factors:

- Topology of the Network:
  - The topology of the network is created by our algorithm. In order to change the topology, we vary attenuation between nodes to break the links or create new links. We measure throughput for different topologies and compare with the performance of the benchmark.

- Size of the Network:
  - Size of the network is measured by the total number of nodes in the network in a given area which essentially indicates the density of network. We evaluate performance of the network with different network densities.

- External and Internal Interference:
  - We perform the experiment in presence of external wireless devices like Bluetooth, Zigbee, microwave oven and other WiFi networks which communicate in 2.4 GHz frequency band and measure the throughput in presence of interference from those sources. For internal interference, we vary the transmission power of the nodes to change the interference range of nodes and measure the effect on throughput of the network.

- Dynamism of the System:
    - Dynamism of the system includes mobility, link disruption and arrival or departure of nodes in the network. We emulate the mobility by physically moving the nodes or by changing the level of attenuation between nodes and observe its effect on the performance.
- Traffic across the Network:
    - For our experiments we keep data traffic requirement of each node as a constant.

Among all these factors, we keep high traffic demand as a static condition for all our experiments and for the comparison benchmark as well. To measure how arrival/departure of nodes affects the throughput of the network, we keep external interference and topology of the network unchanged. For measuring dynamism of the system, we keep the size of the network and external interference constant while changing the topology of the network. Finally, we vary external interference while keeping other two factors constant to analyze how it affects the performance. This way, we evaluate the direct effect of each factor on throughput. Also, we emulate the same effects on the comparison benchmark.

Apart from maximizing the throughput, we want to minimize the latency of communication as well. Also, as we propose a protocol for dual band wireless network, we check the scalability, reliability and adaptability of our solution. Scalability can be verified by increasing the size of the network, reliability can be checked in terms of number of packet drops or Bit Error Rate (BER) and adaptability can be verified while measuring the throughput with dynamism of the system.

Now, for measuring all these metrics we rely on emulating experiments on wireless testbed rather than simulation. The wireless emulator can provide scenarios closer to real world in terms of the factors like interference, dynamism of the network compared to simulation based experiments. Since the emulator can give results which are quite varying because of the fluctuation in the conditions of the systems, we measure the throughput when the system becomes stable after certain duration. We also take average value of significant number of readings to overrule any bias in the reading.

## 5  Solution Description

In this section we provide solution to each subproblem defined in Problem Formulation section. First, we provide a way to measure the quality of the link between a node and an AP when a new node wants to join the network. Next, we provide an algorithm to enable new nodes to join and build multi-hop dual bandwidth network. Then we give a modified version of MAC layer handoff mechanism to cope up with dynamism of the network and provide nodes with a quick way to do handoff. Finally we end this section by providing traffic management and implementation details.

### 5.1  Measuring Link Quality Metric

To measure the link quality, we use received signal strength and available bandwidth of the channel. The received signal strength can be measured by actively scanning the available APs and measuring the RSSI (Received Signal Strength Indicator) values. But the RSSI values alone can only indicate the signal power and its variation with the effect of interference and attenuation. As RSSI does not give a true picture of the load on the AP, we measure the available bandwidth of the channel. So, we use Airtime Link Metric [9] along with the information of average load on AP to measure the link quality metric.

Consider scenarios where the nodes connected to an AP send bursts of traffic in small intervals. In such cases if we use only Airtime Link Metric, the node trying to connect to this AP might get a wrong idea of load on AP. Airtime Link Metric will indicate high load on AP if the measurements are taken when the nodes are sending bursts of data or it will indicate low load if the measurements are taken when nodes are idle. To solve this problem, we maintain average load information on each AP over a period of time and send it to the new node that is trying to connect. The new node will take sum of Airtime Link Metric and average load on AP to decide on which AP to connect to.

Link Quality Metric = Airtime Link Metric + Average load on AP

Lower value of link quality metric indicates better AP.

## 5.2 Algorithm to build a Dual Band Wireless Network

We propose an algorithm to build a dual band (2.4 GHz and 5.8 GHz) wireless network employing WiFi Direct aiming at the improvement of overall throughput of the network. We impose two basic assumptions for our algorithm:

a) Every wireless node is enabled with two WiFi cards, one of which supports 2.4 GHz band and the other supports 5.8 GHz band.
b) All the wireless links are bidirectional.

In order to increase the simultaneous reception and transmission and hence enhance the overall throughput of the network, we propose that a node uses the link of one frequency band to connect to an AP (software or hardware) and hence receives data through that link. While the link of other frequency band be used for transmission of data as an AP to other child nodes.

**Building the Network:**

1. When the first node (say node G) wishes to join the network, it will be the gateway for the mesh network which can connect to the backbone network for internet services. So, the first node connects to the hardware AP using link of one band, say 2.4 GHz band (if the hardware AP is a legacy device which transmits beacons at 2.4 GHz band).
2. Then node G will also transmit beacons using the other link (5.8 GHz in this case) so that it can serve as a software AP to other incoming nodes.
3. When a second node comes in, it hears the beacons transmitted in both 2.4 GHz and 5.8 GHz band links. Then it compares the link quality metric (described in previous section) of both the links and connects to the best available link.
4. This way, the network grows as new nodes join the hardware AP or a software AP and after joining the network, the node itself acts as a software AP using the other frequency band link.

These steps are sufficient to build a dual band wireless network when the network conditions are stable. However, when the network conditions are dynamic, we need to reconfigure the network for which we propose an efficient handoff approach which is described in the next section.

**Efficiency of the Solution:**

- This solution is scalable, as it does not limit the number of nodes in the topology. When a new node comes in, irrespective of the network conditions it follows the same mechanism to join the network.
- This algorithm performs better compared to a single frequency band network because here for any node, transmission and reception can be done concurrently without interference as they are done on different frequency bands. In particular, when the traffic is high and network is dense such that all the non interfering channels of a frequency band are in use, then using dual frequency network significantly improves the throughput by minimizing interference.

## 5.3 Dynamic Handoff

To cope up with the dynamism of the network, we propose an effective handoff mechanism to recon-figure the network. The dynamism of the network can happen because of change in different network conditions like change in topology, breaking of a link, departure of a node from the network, presence and variation of interference, mobility of the nodes etc. In these scenarios, we might have to do MAC layer handoff for a node to maintain the network connection. There are two scenarios to be considered to decide whether handoff should be done or not.

i) Consider the scenario when an existing link becomes weaker and it is no more the best available link for the node but the quality of the link is good enough to carry the required traffic load through it. In this case, employing a handoff would not provide any improvement in throughput. Instead, it increases the overhead and instability, hence affects the performance of the network. So, we will not do handoff in such cases.

ii) The second scenario is when network conditions change severely and the existing link quality goes below the threshold or the link is broken. In that case we need to do handoff to improve the performance. In this situation, there are two cases for doing handoff which we call as hard handoff and soft handoff.

**Hard Handoff:** If the network conditions change only at the edge of the network, we propose a handoff for an orphan node to find the next best link of either frequency bands. If the node decides to connect to an AP on a different frequency band then we call it as hard handoff. Hard handoff for an inner node leads to changing the links to different frequency band for all the nodes in its subnetwork.

**Soft Handoff:** The second case is a soft handoff. This is a kind of optimization we do when an inner node has to do a handoff. As we are using different links for transmission and reception, finding a link of the same frequency as the old link will keep the sub-network of that node unaltered. Thus, it would result in less overhead and quick stabilization of the network. As this handoff shifts to the next best link of the same frequency band, which is not necessarily the overall next best link and it does not force to change any other links in its sub-network, it is called as soft handoff.

**Handoff Mechanism:**

In 802.11 protocol when a node needs to do a handoff, it scans all the channels for an AP and then switches to the channel of the new AP. After switching the channel it authenticates with that AP and connects to it. Even active scanning, which is considered to be better compared to passive scanning in terms of latency, takes hundreds of milli seconds to do a handoff which is not very efficient. So, we employ a modified version of IEEE 802.11 [16] MAC layer handoff mechanism.

The main aim of this paper is to improve the throughput of overall network employing dual band links. Using conventional 802.11 MAC layer handoff will substantially affect the network throughput as the subnetwork connected to the node requiring handoff will be blocked from sending or receiving data when handoff is taking place. To alleviate this problem we are proposing a solution where each node actively scans for APs every T interval on all the channels even when it is connected to AP with good signal strength. Each node maintains a list of APs found during the scanning process in the decreasing order of link quality metric. When the existing link quality goes below a threshold value the node immediately picks the next best available AP from its list and connects to it.

## 5.4 Traffic Management

Traffic management is the term we use to describe how data traffic is handled in the network. The two cases we consider here are, when a node wants to access internet and when a node wants to access a service, provided by some other node in the network, using WiFi direct.

When a node wants to access the Internet, it sends data to the AP it is connected to. The AP acts as a relay to forward the data to its AP on behalf of this node. This process continues till the data packets reach the gateway node which ultimately delivers packets to hardware AP. The hardware AP then routes the packets to backbone network. To receive packets from Internet, the whole network will be acting as a switch. For the first time when hardware AP needs to send a packet to a node, it broadcasts the packet to all its neighboring nodes which in turn forward it to their neighboring nodes. When the packet reaches its destination, the destination sends an acknowledgement. Based on the acknowledgement received, the hardware AP maintains a map of MAC address of destination to that of neighboring node. The whole network, in a way acts as a layer 2 switch.

For the second case where a node wants to access service provided by some other node, it broadcasts a service request with a request id to its neighbors to find out which node is providing that service. Each node that receives this request will broadcast the request again to its neighbors. If the node has already received a request with a particular request id it ignores the request. This process continues till the request reaches a node providing the service. Any node providing the requested service replies to the requesting node. This reply will then be forwarded all the way back to the node that initiated the request. The reply also contains return path from service providing node to the requesting node. After receiving the replies from its neighbors, the requesting node selects the best service providing node based on return path and reply delay. Since all the channels are bidirectional, the node requesting the service can use the return path for accessing the service.

## 5.5 Implementation and Experiments

To implement the algorithm, we modify the MAC layer implementation of WiFi Direct standard in linux kernel. We modify the code of the MAC layer implementation to add active scanning of the channels in two different frequency bands and compare them to select the best channels based on the value of the link quality metric. We also implement the storage aware handoff mechanism in the code to handle the dynamism of the network.

To test our implementation, we use wireless testbed nodes with dual NICs (Network Interface Cards) that support both 2.4 GHz and 5.8 GHz frequency bands. We employ dynamism in the network either by forcefully eliminating some node(s) or adding interference from other sources. We test our algorithm by measuring the aggregated throughput of all the nodes in the network in high traffic demand conditions.

# 6 Performance Evaluation

In this section, we present the experimental results and the performance evaluation of our proposed model in comparison with benchmark. We relied on experiments with real hardware and not on simulation to match with the real-world scenarios. The following parts of this section describes the experimental setup, various test case scenarios we tested, results and analysis and at last, the conclusion.

**Experimental Setup:**

We used four unix machines to build the network topology and conduct experiments. We set up one unix machine as a hardware AP with a single NIC operating in 5.8 GHz, one machine as a software AP with two NICs operating in 2.4 GHz and 5.8 GHz. And, two machines which are used as edge nodes have a single NIC operating in 2.4 GHz. In our set up, software AP is connected to hardware AP in 5.8 GHz and two edge nodes are connected to software AP in 2.4 GHz. We set the bit rate of NICs as 11 Mbps for the experiment. The edge nodes and interface of software AP operating in 2.4 GHz are in one subnet while interface of software AP operating in 5.8 GHz and hardware AP are in a different subnet. We set up routing rules on all the machines to forward the packets in right direction. We also set up server on hardware AP to emulate internet traffic.

**Test Case Scenarios:**

We conducted experiments for two different topologies consisting of 3 nodes and 4 nodes. Our experimental setup is as shown in figure 1 where there is one hardware AP, one software AP and two edge nodes connected to software AP. In the first experiment we took results for the scenario where two nodes connected to software access point are sending data simultaneously while software AP is not sending any traffic. Then, we took results for scenario where two edge nodes and software AP are sending data. Finally, we conducted experiment and took measurements for a different topology consisting of only three nodes (hardware AP, software AP and edge node) where both edge node and software AP are sending data.

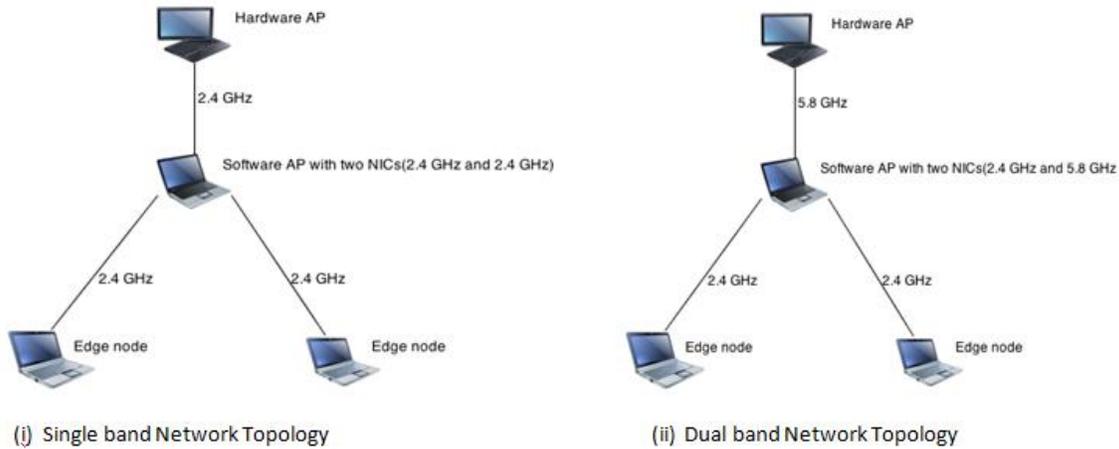

Figure 2: Comparison of throughput of dual band vs single band network

**Results and Analysis:**

The experimental results are shown as below:

Case i) Edge nodes and software AP sending data: In this scenario, the average throughput obtained for dual band network was 3.54Mbps compared with the throughput obtained for single band network which was 1.17Mbps.

Case ii) One edge node and software AP sending data: In this scenario, the average throughput for dual band network was 6.68Mbps while it was 1.628Mpbs for single band network.

Case ii) Only edge nodes sending data: In this scenario, the average throughput was 3.646Mbps for dual band network and .973Mbps for single band network.

# 7 Conclusion

We proposed an efficient way to build a dual band wireless network using WiFi Direct to enhance the throughput and minimize the interference. We performed our experiments with real hardware in presence of interference from other devices. We compared the result with the bench mark where we used two 2.4 GHz cards on each node. Based on the experimental results we found that using dual band (2.4 Ghz and 5.8 GHz) WiFi cards improves the throughput when the number of node increases and traffic is more. With our experimental setup, we build a dual band network, where we can test up to 2-hop wireless communication. But as we stated, our solution is scalable and works for larger topology with more number of nodes.

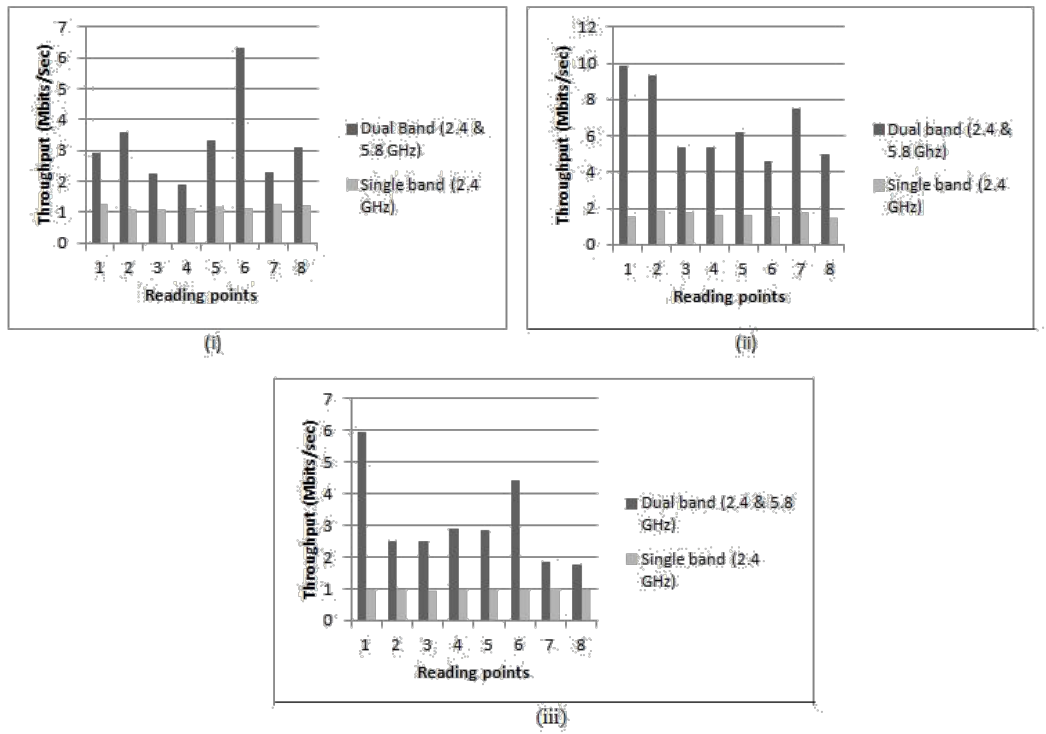

Figure 3: Comparison of throughput of dual band vs single band network

# References

[1] Ian F. Akyildiz, Xudong Wang, and Weilin Wang. Wireless mesh networks: a survey. Computer Networks Volume 47, Issue 4, pages 445–487, 2005.

[2] Mansoor Alicherry and Randeep Bhatia. Joint Channel Assignment and Routing for Throughput Optimization in Multi-radio Wireless Mesh Networks. MobiCom '05, pages 58–72, 2005.

[3] Wi-Fi Alliance. Wi-fi CERTIFIED Wi-Fi Direct Frequently Asked Questions. Wi-Fi Alliance, pages 1–4, 2009.

[4] Wi-Fi Alliance. Wi-fi certified wi-fi direct personal, portable wi-fi technology. Wi-Fi Alliance, pages 1–14, 2010.

[5] Wi-Fi Alliance. Wi-fi CERTIFIED TDLS: Easy-to-use, security-protected direct links to improve performance of Wi-Fi devices. Wi-Fi Alliance, pages 1–14, 2012.

[6] Johannes Berg. Wi-Fi Peer-to-Peer on Linux. Intel Corporation, pages 1–29, 2010.

[7] Daniel Borkmann and Tobias Kalbitz. LinGrok - Linux Kernel Cross Reference (LXR). http://lingrok.org/xref/linux-net-next/drivers/net/wireless, 2012.

[8] Andrew Brzezinski, Gil Zussman, and Eytan Modiano. Enabling Distributed Throughput Maxi-mization in Wireless Mesh Networks - A Partitioning Approach. MobiCom '06, 2006.

[9] Joseph D. Camp and Edward W. Knightly. The IEEE 802.11s Extended Service Set Mesh Networking Standard. IEEE Communication Magazine Vol. 48, Issue 8, 2008.

[10] Teemu Karkkainen, Mikko Pitkanen, and Jorg Ott. Enabling Ad-hoc-style communication in public WLAN hot-spots. Proceedings of the seventh ACM International Workshop on Challenged Networks, pages 31–38, 2012.

[11] Ki-Dong Lee. Fair Allocation of Subcarrier and Power in an OFDMA Wireless Mesh Network. IEEE Journal on Selected Areas in Communications, 2006.

[12] Carlo Parata, Vincenzo Scarpa, and Gabriella Convertino. Flex-WiFi: a mixed infrastructure and ad-hoc IEEE 802.11 network for data traffic in a home environment. World of Wireless, Mobile and Multimedia Networks, 2007. (WOWMOM), pages 1–6, 2007.

[13] Ashish Raniwala and Tzi-cker Chiueh. Architecture and Algorithms for an IEEE 802.11-Based Multi-Channel Wireless Mesh Network. INFOCOM 2005. 24th Annual Joint Conference of the IEEE Computer and Communications Societies. Proceedings IEEE, pages 2223–2234, 2005.

[14] Kiyoshi Ueda, Sumio Miyazaki, Tetsuya Iwata, Hiroyuki Nakamura, and Hiroshi Sunaga. Peer-to-Peer Network Topology Control within a Mobile Ad-hoc Network. APCC, pages 1–5, 2003.

[15] Yi Yang, Mahesh Marina, and Rajive Bagrodia. Evaluation of multihop relaying for robust vehicular internet access. 2007 Mobile Networking for Vehicular Environments, pages 19–24, 2007.

[16] Zhenxia Zhang, Richard W. Pazzi, and Azzedine Boukerche. A Fast MAC Layer Handoff Protocol for WiFi-Based Wireless Networks. WLN, pages 1–7, 2010.